# Datta-Das type spin-field effect transistor in non-ballistic regime

Munekazu Ohno[1], Kanji Yoh[1,2]

[1]*Research Center for Integrated Quantum Electronics, Hokkaido University, Sapporo, 060-8628, Japan*
[2]*CREST-JST, Kawaguchi, Saitama 332-0022, Japan*

June 25, 2007

**Abstract**

It is revealed that in spin helix state of (001) quantum well system, strong suppression of D'yakonov-Perel' spin relaxation process occurs by an interplay between Rashba and Dresselhaus couplings over a wide range of Rashba coupling strength. Contrary to common belief in early works, this leads to the finding that Datta-Das type spin-field effect transistor is actually applicable to more realistic non-ballistic transport regime in two dimensional electron gas system.

*Keywords*: Spin transport, Spin relaxation, Spontronics, Spin-orbit coupling, Rashba effect, Dresselhaus effect

**1. Introduction**

Understanding and controlling of carrier spin transport phenomena in a semiconductor are among the most important and interesting issues in the emerging field of spintronics, being of critical relevance to realization of spin-related devices which provides high performance and novel functionality and may furthermore leads to a quantum computing [1]. A seminal concept of spin-field effect transistor (spin-FET) has been proposed by Datta and Das [2] in the light of controllability of spin precession motion of electron through the Rashba spin orbit coupling effect [3]. It is currently common belief that the Datta-Das type spin-FET can operate only in regime of ballistic transport or (quasi) one-dimensional transport, since spin relaxation mainly due to D'yakonov-Perel'(DP) mechanism [4] causes a loss of information fed by spin coherence [5, 6]. In this letter, we show that the Datta-Das type device is actually applicable to non-ballistic regime in two dimensional electron gas (2DEG) system, which solves the problems hampering practical realization of spin-FET and will provide a substantial progress in development of spin-related device utilizing spin precession motion.

In (001) asymmetric quantum well (QW) system with spatial coordinate of $x//[100]$ and $y//[010]$, which is our concern, bulk-inversion asymmetry (BIA) and structure inversion asymmetry (SIA) leads to the Dresselhaus effective magnetic field, $\mathbf{\Omega}_D(\mathbf{k})=2\beta\hbar^{-1}(-k_x, k_y, 0)$ [7], and the Rashba effective magnetic field, $\mathbf{\Omega}_R(\mathbf{k})=2\alpha\hbar^{-1}(k_y, -k_x, 0)$ [3], respectively, where $k_i(i=x, y)$ is electron wave vector, $\beta$ is material dependent constant and $\alpha$ is the Rashba constant depending on gate voltage. It has been demonstrated in several works [8-10] that when $\alpha=\beta$ is realized, the effective magnetic field is oriented in [110] axis irrespective of $\mathbf{k}$ and hence the spin along [110] axis undergoes neither spin precession nor relaxation. Importantly, the existence of $SU(2)$ symmetry in spin-orbit Hamiltonian at $\alpha=\beta$ has been recently discovered in Ref. [11], claiming that at $\alpha=\beta$, the spin precession motion of electron traveling through $[1\bar{1}0]$ spatial coordinates does not suffer from the DP spin relaxation process and the spatially coherent spin oscillation pattern emerges, which is called Persistent Spin Helix (PSH) state. It should be pointed out that the early studies on suppression of spin relaxation have been centered around this special point of α=β and despite its importance, robustness of spin helix state against relaxation over a wide range $\alpha$ value has not been understood. In this study, this robustness is scrutinized and clarified by means of computational experiment, i.e., semi-classical Monte-Carlo (MC) simulation. It is demonstrated that the interplay between Rashba and Dresselhaus couplings strongly suppresses the spin relaxation process in spin helix state over a wide range of Rashba coupling strength and this leads to the finding that contrary to common belief, the Datta-Das device can successfully operate in non-ballistic regime of 2DEG system.

## 2. Semi-classical Monte Carlo approach for D'yakonov-Perel' process

The spin relaxation during carrier transport takes place by several mechanisms [4, 12-14]. At room temperature, electron spin relaxation due to the DP mechanism is the most dominant in QW system grown on (001) substrate. The detail of the semi-classical MC simulation of DP process has been discussed in Ref. [6, 15-17] and, hence, only the essential points are briefly summarized in the following. The MC simulation is applied to the DP process at 300 K in (001) $In_{0.81}Ga_{0.19}As$ QW of 80 Å well width inserted between $In_{0.53}Ga_{0.47}As$ subchannel and $Al_{0.48}In_{0.52}As$ barrier layer, which is one of the candidate structures for realizing spin-FET [18]. We performed the ensemble MC simulation of $4\times10^4$ electrons in 2DEG system, taking into account elastic scatterings originating from polar optical phonon, acoustic phonon, remote impurity and alloying effect. The details of these scattering rates for the 2DEG system are found in Ref. [19]. In the simulation, the gate length is set to be $L$ μm and the gate width is assumed be to infinite. Note that the condition of one-dimensional transport is not introduced in the present analysis. The spatial distribution of electron over [110] coordinate is assumed to be uniform and our concern is placed on observation along [1$\bar{1}$0] spatial coordinate between the drain and source. The carrier mobility is calculated to be $9.70×10^3$ cm$^2$/(V·s) in the present MC simulation, which is quite comparable to $9.75×10^3$ cm$^2$/(V·s) experimentally measured for the same structure [18].

The strengths of Dresselhaus and Rashba effects are determined by the constants $\beta$ and $\alpha$, respectively. In all the calculations, the constant $\beta$ is assumed to be $9.2×10^{-12}$ eV·m which is compositionally averaged value of $\beta$ in InAs and GaAs [20]. As for the Rashba constant $\alpha$, we vary this value systematically, using a ratio of $\alpha/\beta$ which is called RD-ratio here. Our focus is placed on the range of RD-ratio between 0 and 3 and this corresponds to the variation of $\alpha$ between 0 and $27.6×10^{-12}$ eV·m which is in reasonable range of $\alpha$ value in InAs-rich channel layer [21, 22]. It is to be noted here that when $\alpha=\beta$ is realized, the effective magnetic field is oriented in [110] axis irrespective of **k**. Hence if the injected spin is parallel to [110] axis, the spin experiences neither spin precession nor relaxation and this is the essential requirement for a non-ballistic spin-FET proposed in Ref. [8] in which spin precession motion is completely suppressed. On the other hand, in the present simulation, the injected spin is set to be parallel to [1$\bar{1}$0] axis and hence the spin precession is involved in electron traveling through the channel layer. In the simulation, when the electron reaches the drain, it is removed from the channel and is again injected at the source. By repeating this process, the spin polarization is averaged over some period at each [1$\bar{1}$0] spatial coordinate and this averaged value is denoted as <**S**>$_T$.

**3. Results and Discussion**

Shown in Fig. 1(a) are the spatial profiles of spin polarization vector calculated at $\alpha/\beta$=0.8(upper), 1.0(middle) and 1.5(lower) where <$S_+$>$_T$, <$S_-$>$_T$ and <$S_z$>$_T$ describe the averaged spin polarization components of [110], [1$\bar{1}$0] and [001] axes, respectively. The gate length is set to be 2 μm and the in-plane field, $E_{ds}$, is 0.5 kV/cm. Figure 1(b) represents the spin relaxation processes calculated with the same condition but without Dresselhaus term, which is equivalent to the original Datta-Das proposal. The spatial oscillation of spin polarization is observed in all the cases. As shown in Fig. 1(b), during the spin transport with only the Rashba coupling effect, the coherence of spin polarization is completely lost at $L$=2 μm, because of strong spin relaxation. In the system with the Rashba and Dresselhaus effects (Fig. 1(a)), on the other hand, the coherence of spin polarization persists even at $L=2$ μm at all three values of RD-ratio. Especially, no spin relaxation occurs at $\alpha/\beta=1.0$ and this spatial pattern is exactly equivalent to PSH state [11, 17]. The precession angle of spin during a certain time period $\Delta\tau$ can be expressed as $\theta=|\mathbf{\Omega}_{eff}(\mathbf{k})|\Delta\tau=2m^*r\hbar^{-2}(\alpha^2+\beta^2-2\alpha\beta\sin2\theta)^{1/2}$ where $\mathbf{\Omega}_{eff}(\mathbf{k})=\mathbf{\Omega}_R(\mathbf{k})+\mathbf{\Omega}_D(\mathbf{k})$, $m^*$ is the electron effective mass, $r=\hbar\cdot(k_x^2+k_y^2)^{1/2}\cdot\Delta\tau/m^*$ is the magnitude of moving distance of electron from the source and $\theta$ is the angle of wave vector with respect to $x$ axis. When the relation $\alpha=\beta$ is realized, the effective magnetic field is oriented in [110] axis irrespective of $\mathbf{k}$ and the precession angle is reduced to $\theta=4\alpha m^*\hbar^{-2}r_{1\bar{1}0}$ where $r_{1\bar{1}0}=r\cos(\theta+\pi/4)$. The precession angle is thus proportional to the displacement of electron along $[1\bar{1}0]$ spatial coordinate, which is the origin of the PSH states [17, 23]. The effective magnetic field parallel to [110] does not cause the spin relaxation during electron transport along $[1\bar{1}0]$ direction. Therefore, it can be stated that the spin relaxation at $\alpha\neq\beta$ is entirely ascribable to the existence of effective magnetic field of which orientation deviates from [110] axis. Although the Rashba effective field is always perpendicular to $\mathbf{k}$, the interplay between the Rashba and Dresselhaus effects makes the effective field parallel to [110] dominant in the vicinity of $\alpha/\beta=1.0$, thus suppressing the spin relaxation as observed in Fig. 1. This fact is the key to the successful operation of Datta-Das device in non-ballistic regime.

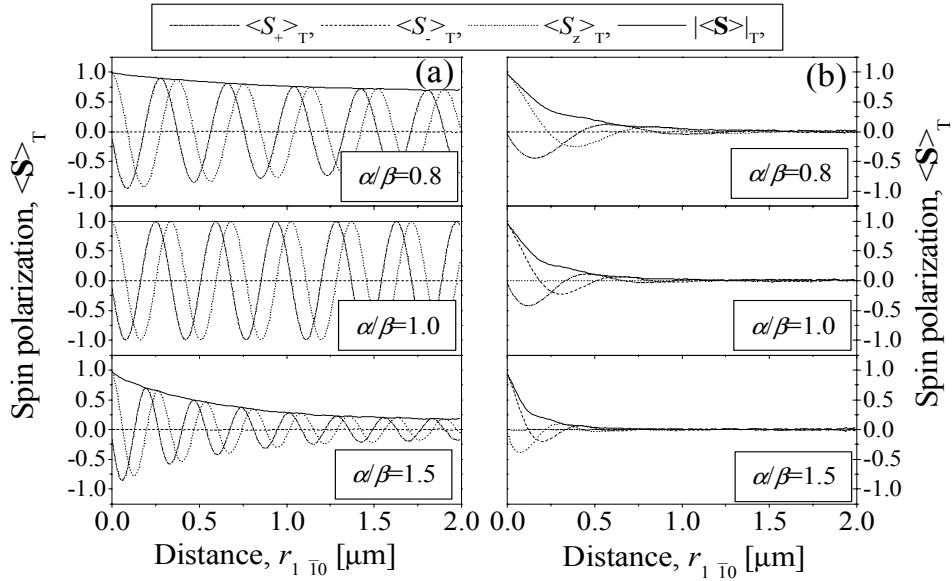

Figure 1. (a) Spatial profiles of $<\mathbf{S}>_T$ at $\alpha/\beta = 0.8$ (upper), 1.0 (middle) and 1.5 (lower). $|<\mathbf{S}>|_T$ is given as $|<\mathbf{S}>|_T = (<S_+>_T^2 + <S_->_T^2 + <S_z>_T^2)^{1/2}$. The in-plane field, $E_{ds}$, is 0.5 kV/cm. (b) Spatial profiles of spin polarization calculated with the same condition as the ones in (a) but without the Dresselhaus coupling.

Figure 2 represent the spin polarization component $<S_->_T$ plotted with respect to $[1\bar{1}0]$ spatial coordinate and RD-ratio calculated with both the Rashba and Dresselhaus couplings. Our MC simulation demonstrated that the coherence of spin polarization at a fixed spatial coordinate increases with in-plane field as is similar to the discussion of "upstream" spin diffusion length in a drift-diffusion approach [24], while a period of spatial oscillation is independent of in-plane field. With increasing in-plane field, the number of electrons contributing to current transport increases at $\mathbf{k}//[1\bar{1}0]$ where $\mathbf{\Omega}_{eff}(\mathbf{k})$ is always oriented to [110] irrespective of RD-ratio, thus suppressing spin relaxation. The result in Fig. 2 corresponds to the one at $E_{ds}$=3.0 kV/cm where the oscillation pattern can be clearly seen. The substantial spin relaxation is not involved in the vicinity of $\alpha/\beta$=1.0. It is quite important to note here that an oscillation can be seen at a fixed spatial coordinate by changing RD-ratio, as the current oscillation due to variation of gate voltage in the Datta-Das type spin-FET. Such an oscillation of $<S_->_T$ at $L$=2 μm is demonstrated by solid line in Fig. 3(a) where the magnitude of spin polarization vector, $|<\mathbf{S}>|_T$, is also given by dashed line. In the present device, the "1" and "0" states correspond to upper and lower peaks in this oscillation, respectively. The period of oscillation, $\Delta\alpha/\beta$, which does not obviously change over this range of RD-ratio, is about 0.15 and this is entirely associated with the variation of gate voltage required to switch "1" and "0" states. Importantly, the amplitude of oscillation denoted as $\Delta<S_->_T$ in the vicinity of $\alpha/\beta$=1.0 is close to maximum value, 2.0, realized without spin relaxation. It is therefore comprehended that the Datta-Das type device can successfully operate in non-ballistic regime. In Fig. 3(a), moreover, the calculated result at $E_{ds}$=0.5 kV/cm is represented by dotted line. The peak positions are identical with the ones at $E_{ds}$=3.0 kV/cm. Although the amplitude of oscillation becomes small and vanishes at low and high values of $\alpha/\beta$, $\Delta<S_->_T$ in the vicinity of $\alpha/\beta$=1.0 is considered as sufficiently large to allow the successful operation of device.

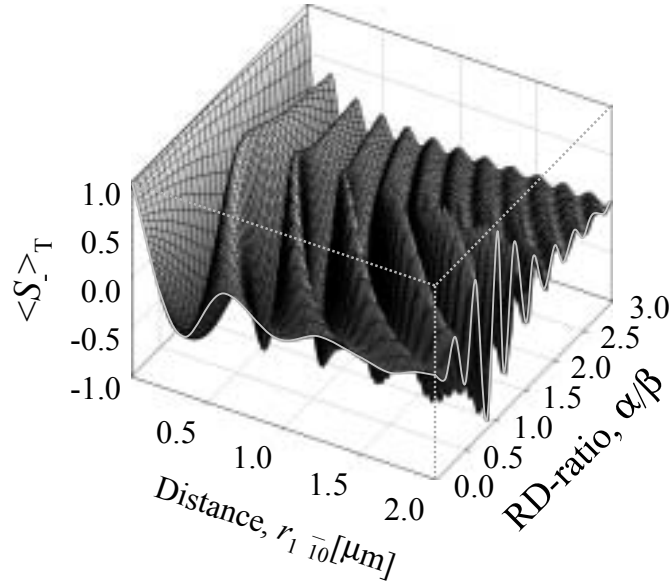

Figure 2. Spin polarization component, $<S_->_T$ with respect to $r_{1\bar{1}0}$ and $\alpha/\beta$ calculated at $E_{ds}$=3.0 kV/cm.

It is generally expected that $\Delta<S_->_T$ should be large in a small gate length, since the coherence of spin polarization persists. In Fig. 3(b), the oscillations of spin at $L$=2 μm is compared with the one at $L$=0.5 μm. As expected, the magnitude $|<\mathbf{S}>|_T$ at $L$=0.5 μm is always larger than the one at $L$=2 μm. However, $\Delta<S_->_T$ becomes small at $L$=0.5 μm. From the equation of precession angle $\theta=|\mathbf{\Omega}_{eff}(\mathbf{k})|\Delta\tau$, $\Delta\alpha/\beta$ required for π rotation at $\alpha=\beta$ and $r_{1\bar{1}0}=L$ is obtained as $\Delta\alpha/\beta=\pi\hbar^2/(2m^*\beta L)$. Hence, $\Delta\alpha/\beta$ increases with decreasing gate length $L$, which results in the peak positions for "1" and/or "0" states deviated far from $\alpha/\beta$=1.0, leading to the decrement of $\Delta<S_->_T$, as seen in Fig. 3(b). In other wards, the long gate length results in small $\Delta\alpha/\beta$ and thus relatively large $\Delta<S_->_T$ because both "1" and "0" states can be located in the vicinity of $\alpha=\beta$.

It should be mentioned that although our focus in this paper has been directed to the specific structure of $In_{0.81}Ga_{0.19}As$ QW, the main conclusion of this study does not lose the generality, that is, successful operation of non-ballistic Datta-Das device is considered possible in the other structures. Our calculations showed that the difference in carrier mobility associated with gate voltage does not yield substantial change in spatial profiles of spin within the focused range of RD-ratio, which indicates that detail of contribution of different scattering events and their frequencies do not play a deterministic role in successful operation of this non-ballistic device. Moreover, it has been experimentally observed in Ref. [25] that the value of RD-ratio focused in the present study can be realized in several types of structures like InAs/InAlAs and GaAs/AlGaAs. Furthermore, it would be worthwhile to mention the following points neglected in this work. In the present calculations, we ignored the cubic term in Dresselhaus model. In existence of the cubic term, the PSH state suffers from the spin relaxation and the spin diffusion length becomes finite even at $\alpha=\beta$ [11]. However, it is expected that such a finite value of spin diffusion length is still sufficiently large to allow a successful operation of the present device [17]. Most importantly, details of injection and detection of spin polarized current were not addressed in the present work. In this regard, we would like to refer to experimental works [26, 27] where successful injection of spin polarized current into InAs channel has been demonstrated.

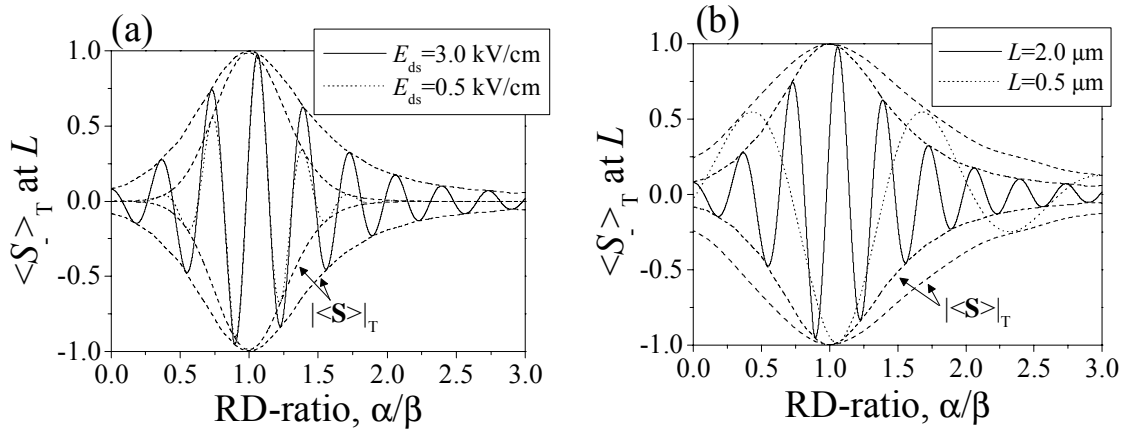

Figure 3. (a) Spin polarization component $<S_->_T$ at $L$=2 μm calculated at different $E_{ds}$. (b) Spin polarization component $<S_->_T$ at $E_{ds}$=3.0 kV/cm calculated at different $L$.

In summary, we performed Monte-Carlo analysis on robustness of spin helix state against spin relaxation. It was demonstrated that the interplay between Rashba and Dresselhaus couplings strongly suppresses the spin relaxation process in spin helix state over a wide range of $\alpha/\beta$ value, compared to the process with only the Rashba coupling effect. Contrary to common belief, our outcomes suggest that the Datta-Das device can successfully operate in non-ballistic regime of 2DEG system. This fact will contribute to substantial progress in practical realization of the spin-related devices, especially spin field-effect transistors.

**Acknowledgment**

This work was partly supported by Grant-in-Aid for Scientific Research from the Japanese Ministry of Education, Culture, Sports, Science and Technology.